# Gate-tunable single-photon electroluminescence of color centers in silicon carbide

Igor A. Khramtsov and Dmitry Yu. Fedyanin*

*Laboratory of Nanooptics and Plasmonics, Moscow Institute of Physics and Technology, Dolgoprudny 141700, Russian Federation.*

E-mail: dmitry.fedyanin@phystech.edu

Electrically driven single-photon sources are essential for building compact, scalable and energy-efficient quantum information devices. Recently, color centers in SiC emerged as promising candidates for such nonclassical light sources. However, very little is known about how to control, dynamically tune and switch their single-photon electroluminescence (SPEL), which is required for efficient generation of single photons on demand. Here, we propose and theoretically demonstrate a concept of a gate-tunable single-photon emitting diode, which allows not only to dynamically tune the SPEL rate in the range from 0.6 to 40 Mcps but also to switch it on and off in only 0.2 ns.

Novel devices exploiting the quantum nature of light are in great demand for the emerging quantum information technologies, such as quantum cryptography and optical quantum computing. Most of these devices exploit nonclassical states of light that can be produced only by a single-photon source - an optical source that generates a train of optical pulses so that each pulse contains strictly one photon. In the past two decades, many efforts have been made to obtain single-photon generation under electrical excitation, which is highly desirable for practical applications.[1] Different material platforms were proposed, such as epitaxial and colloidal quantum dots,[2,3] single molecules, and nanotubes.[4] However, all these approaches have many drawbacks related to their stability, operating temperature, and fabrication techniques, which greatly limit the range of possible applications. Recently, color centers in diamond and other wide-bandgap semiconductors emerged as promising candidates for practical single-photon emitters.[5,6] Many of these point defects in the lattice demonstrate bright and stable single-photon luminescence with a narrow emission spectrum at room and higher temperatures.[7] At the same time, color centers can be excited electrically,[8–12] which allows their integration into practical quantum information devices. Among all wide-bandgap semiconductors that can host color centers, silicon carbide stands out as the only material with a well-developed CMOS-compatible fabrication process flow.[13] Recent studies have revealed that under electrical excitation, color centers in SiC are brighter than their counterparts in other wide-bandgap semiconductors.[10,14,15] Moreover, they can emit photons at telecommunication wavelengths,[16] which is crucially important for practical implementation of secure communication lines based on quantum cryptography. At the same time, the latest findings show that the process of single-photon electroluminescence (SPEL) of color centers is remarkably different from the electroluminescence of quantum dots and single molecules.[15] Therefore, further research aimed at understanding how to control the SPEL of color centers, dynamically tune and quickly switch it on and off is needed for the development of efficient and controllable single-photon sources based on silicon carbide.

In this Letter, we investigate the possibility to electrically control single-photon electroluminescence of a color center in a silicon carbide p-i-n diode and present a concept of a gate-tunable single-photon emitting diode (GT-SPED). Using a comprehensive computational approach, we demonstrate not only how to precisely tune the SPEL rate of the color center but also how to switch it on and off almost immediately using the gate electrode. We find the switching time to be less than 0.2 ns, which is more than two orders of magnitude faster than the *RC* time constant of the forward biased silicon carbide p-i-n diode.

Figure 1(a) shows a schematic of the 4H-SiC gate-tunable single-photon emitting diode (GT-SPED). Despite that at first glance, it looks very similar to the metal-oxide-semiconductor field-effect transistor (MOSFET), it is entirely different. Unlike the unipolar MOSFET, the GT-SPED is a bipolar device based on a lateral p-i-n diode, which is essential, since for the single-photon electroluminescence (SPEL) one needs both electrons and holes.[15,17] In MOSFETs, electrons (or holes) flow from source to drain in a relatively narrow n-type (or p-type) channel.[18] In the GT-SPED, electrons flow from the n-type contact to the p-type contact and holes flow from the p-type contact the to n-type contact. In addition, there is no channel, i.e., even a semi-insulating substrate is actively involved in the electron and hole transport.

The GT-SPED can be easily fabricated on a semi-insulating 4H-SiC substrate by implanting aluminum and nitrogen atoms and thus forming the p-type and n-type regions, respectively.[19] After that, a color center can be created in the i-region of the diode right below the SiC surface using either focused ion beam implantation[20] or annealing,[10,21] or electron irradiation,[22] or another technique. Depending on the method used, a thin defective layer can appear on top of the sample, but it can be etched away[23,24] to reduce the number of defects in the vicinity of the color center of interest. Etching also allows to decrease the distance from the color center to the top surface of the sample, i.e., to the ITO gate electrode (see Fig. 1(a)), which is beneficial for the improvement of tunability of the GT-SPED, as discussed below. A 50-nm-thick and 1-μm-wide ITO gate is used to control the luminescence of the color center in the i-region of the diode. The ITO gate is optically transparent absorbing less than 1% of photons emitted by the color center.[25] A 20-nm-thick $SiO_2$ layer required to isolate the ITO gate from the semiconductor can be either grown thermally or deposited on the SiC surface.[13] This thickness is high enough to avoid the influence of the ITO gate on the quantum efficiency of the color center.[25] The doping profile is shown in Fig. 1(a). The dopant concentrations are high enough to form low-resistance ohmic contacts to the diode.[26] Other parameters of the GT-SPED are listed in Supplementary Information.

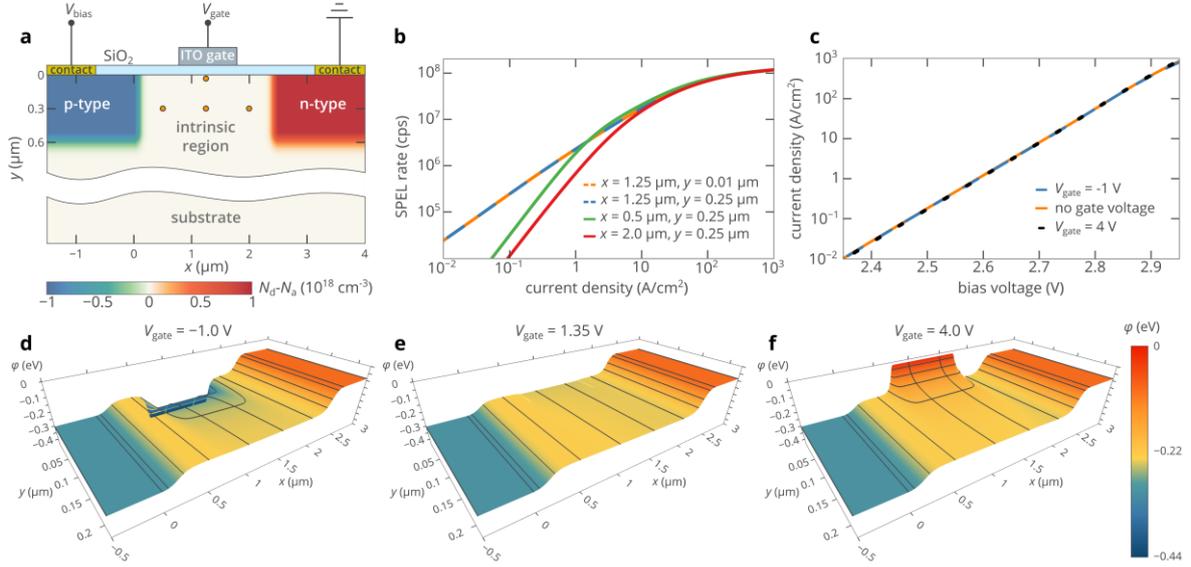

**Fig. 1.** (a) Schematic illustration of the central part of the 4H-SiC GT-SPED. The p-type and n-type contacts are shown not to scale (see Fig. S1 in Supplementary Information). The 2D map shows the doping profile. Four orange dots in the i-region illustrate the positions of the color center used in panel (b). (b) Dependence of the SPEL rate on the current density at the p-type (or n-type) contact for four different positions of the $Si_C$-SF center shown in panel (a) when no voltage is applied to the ITO gate. To obtain the count rate at the photodetector, the SPEL rate should be multiplied by the collection efficiency. (c) Current-voltage characteristics of the GT-SPED at different gate voltages. (d-f) Electrostatic potential distribution in the devices at three gate voltages: -1.0 V (d), 1.36 V (e), and 4.0 V (f).

The SPEL process of color centers is governed by the electron and hole capture processes, as recently revealed by Fedyanin et al.[15,17] Therefore, the SPEL rate of the GT-SPED is a function of the lifetimes of the excited states of the color center and the electron and hole densities in its vicinity. Following the methodology of Ref. 15, we self-consistently simulate the GT-SPED using the nextnano++ software (nextnano GmbH, Munich, Germany) and, using these results, find the SPEL rate at different bias and gate voltages. Although 4H-SiC can host a number of different color centers,[27] here we consider the silicon antisite near stacking faults ($Si_C$-SF)[10] (see Table S2 in Supplementary Information) since it is the only color center in 4H-SiC for which the mechanism of electrical excitation is already understood, and the theoretical predictions can accurately reproduce the experimental measurements.[15] However, we note that our results can be easily extended to other color

centers.[15]

Figure 1(b) shows the SPEL rate of the $Si_C$-SF center in the i-region of the GT-SPED when no voltage is applied to the ITO gate. At current densities above 10 A/cm$^2$, the electrons and holes are distributed nearly uniformly across the i-region of the diode, and therefore, the SPEL rate almost does not depend on the position of the color center in the i-region of the p-i-n diode. At $J$=30 A/cm$^2$ ($V_{bias}$=2.77 V), the electron and hole densities in the i-region are equal to $1.6\times10^{15}$ cm$^{-3}$. The SPEL rate is as high as 38 Mcps, which is only four times lower than the maximum possible SPEL rate of the $Si_C$-SF center in bulk 4H-SiC $\tau_r \tau_{nr}/(\tau_s+\tau_{nr})$ = 145 Mcps. Here, $\tau_r$=3.6 ns and $\tau_{nr}$=36 ns are the radiative and nonradiative lifetimes of the excited state, and $\tau_s$=33 ns is the lifetime of the shelving state of the $Si_C$-SF center.[10,15]

By applying a voltage to the ITO gate, we can change the electron and hole densities in 4H-SiC below the gate. While in MOSFETs, the gate voltage increases the density of either electrons or holes, depending on the type of the channel, in the GT-SPED it affects both electrons and holes. Figure 1(d-f) shows the potential distribution at three gate voltages. At $V_{gate}$=1.35 V (Fig. 1(e)), the gate does not have any influence on electrons and holes in silicon carbide, which can also be seen in Figs. 2(a) and 2(b): the electron and hole distributions coincide with those obtained when no voltage is applied to the gate. At $V_{gate}$=4.0 V (Fig. 1(f)), a high potential hill is formed below the gate, which depletes the holes underneath the gate by two orders of magnitude (see Fig. 2(b)). Figure 3 shows that at $V_{gate}$=4.0 V, no holes are delivered to this region below the gate. At the same time, this potential hill is a potential well for electrons, which are accumulated there (Fig. 2(a)). Thus, a 100-nm-thick region with a high density of electrons and a low density of holes is formed below the gate. It is important to note that this thickness is determined by the electron and hole densities in the i-region of the GT-SPED, which are approximately equal to each other. The only possibility to increase the thickness of the region with a high electron density is to decrease the bias voltage and, consequently, reduce the SPEL rate, which is not reasonable for practical applications.

Since the SPEL rate of the color center is given by[15]

$$R = \left[\left(\frac{1}{c_n n}+\frac{1}{c_p p}\right)\left(1+\frac{\tau_r}{\tau_{nr}}\right)+\tau_r\left(1+\frac{\tau_s}{\tau_{nr}}\right)\right]^{-1} \quad (1)$$

where $n$ and $p$ are the electron and hole densities in the vicinity of the color center, and $c_n$ and $c_p$ are the electron and hole capture constants (see Table S2 in Supplementary Information), it is determined by $\min(c_n n, c_p p)$. Therefore, the SPEL rate of the color center located below the gate decreases as the gate voltage increases (Fig. 2(c)). Similarly, by decreasing $V_{gate}$ below 1.35 V, one depletes the electrons and accumulates holes, decreasing the SPEL rate (Fig. 2(c)).

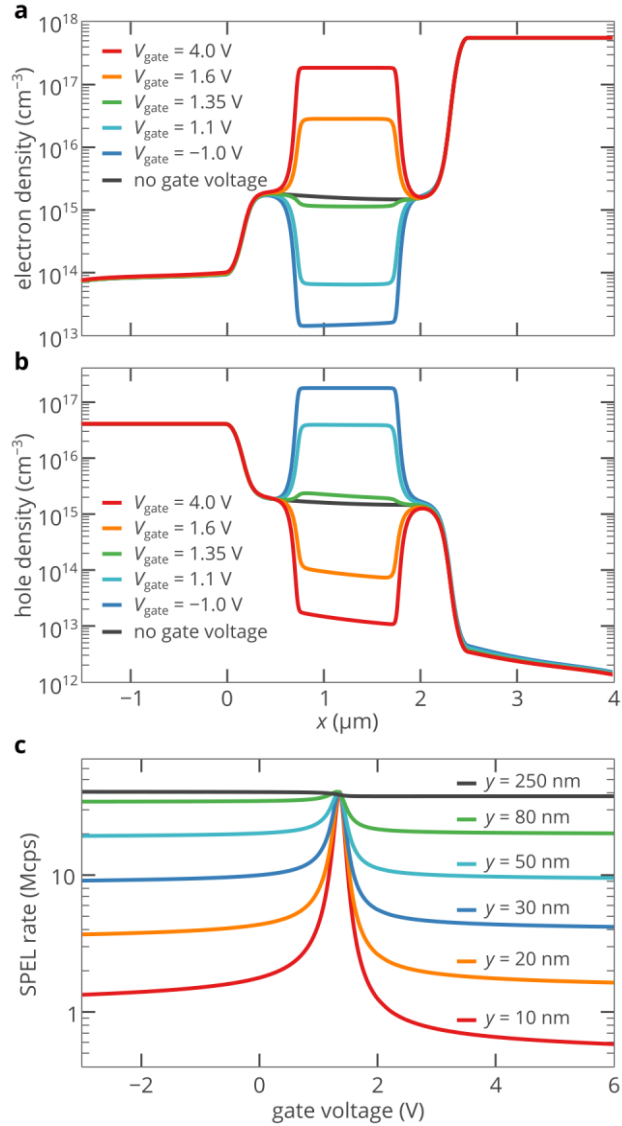

**Fig. 2.** (a,b) Electron (a) and hole (b) distributions along the *x*-axis at *y*=10 nm for different gate voltages. (c) Dependence of the SPEL rate on the gate voltage for the $Si_C$-SF center located underneath the gate electrode at *x*=1.25 μm for different distances between the color center and the $SiO_2$/SiC interface. To obtain the count rate at the photodetector, the SPEL

rate should be multiplied by the collection efficiency.

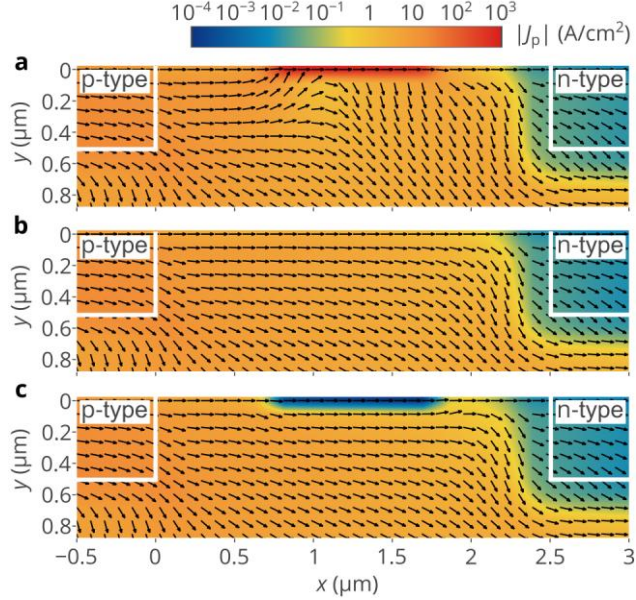

**Fig. 3.** 2D map of the hole current density $|J_p|$ in the central region of the GT-SPED at $V_{bias}$=2.77 V for three gate voltage: -1 V (a), 1.35 V (b), and 4 V (c). Arrows show the direction of the hole flow.

Figure 2c shows that at $V_{bias}$=2.77 V, the SPEL rate of the $Si_C$-SF center located at $y$=10 nm decreases from 40 Mcps to 0.66 Mcps as $V_{gate}$ increases from 1.35 V to 4 V. Thus, the brightness of the $Si_C$-SF center can be easily tuned by the gate voltage, which is especially important for the development of large-scale circuits, which should contain a number of single-photon sources with absolutely the same properties. We should mention that the tuning has little effect on the quantum correlation between emitted photons (Fig. (4)). First, we operate not far from the maximum SPEL rate of the $Si_C$-SF center, where the second-order autocorrelation function $g^{(2)}(\tau)$ is not very sensitive to the carrier capture rates.[15,28] Second, as $V_{gate}$ increases, decreasing the hole density in the vicinity of the color center, the electron density in the vicinity of the color center increases (see Figs. 2(a) and 2(b)), while the $g^{(2)}$ function is determined only by the fastest process among the electron capture and hole capture processes.[28] It is also important to note that the gate voltage almost does not affect the current-voltage characteristics and consequently the heating power given by $JV$ (Fig. 1(c)), which allows to stabilize the device temperature and consequently the emission properties of the color center. Figure 1(d-f) shows that the gate changes the electrostatic potential only in a relatively thin region in silicon carbide below the gate. Thus, only color centers located less than 50 nm below the SiO$_2$/SiC interface can be efficiently controlled by the gate, which is shown in Fig. 2(c). However, this issue is not a serious problem for modern fabrication techniques, which allow to create color centers at a depth of less than 20 nm with high lateral resolution (e.g., see Ref. 20). The SPEL contrast ratio for near-surface defects is as high as 25 for the Si$_C$-SF center at $y\approx 20$ nm and 60 for the Si$_C$-SF center at $y\approx 10$ nm.

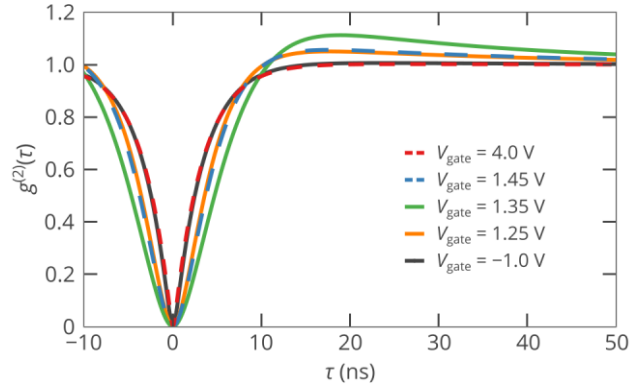

**Fig. 4.** $g^{(2)}$ functions for the Si$_C$-SF center located below the gate at a distance of 20 nm from the SiO$_2$/SiC interface for different gate voltages at $V_{bias}$=2.77 V.

The high contrast between the SPEL rates at $V_{gate}$=1.35 V and $V_{gate}$=4 V gives the possibility to switch the SPEL on and off simply applying a voltage to the gate at a fixed bias voltage. It is remarkable that the GT-SPED responds to the gate voltage change much faster than to the bias change, which allows to switch the SPEL rate at very high speed. The GT-SPED is a bipolar device. Therefore, the capacitance per unit device width $C_{n-p}$ between the n-type and p-type contacts of the forward-biased GT-SPED is rather high due to the charge carrier diffusion capacitance.[18] Using the thermodynamic definition of capacitance[29] and the distributions obtained in the numerical simulations of the electron and hole transport in the steady state, we find $C_{p-n}$ to be about 3 nF/cm at $V_{bias}$=2.77 V. The resistance in the transient process can be estimated from the current-voltage characteristics $V_{bias}(I)$ as $R_{p-n} = dV_{bias}/dI = 10$ Ωcm, which gives a characteristic time of $R_{p-n}C_{p-n}$=30 ns. This relatively low characteristic time is comparable with the inverse electron and hole capture rates, but we can switch the GT-SPED much faster by applying voltage to the gate at a fixed bias voltage. The capacitance of the ITO/SiO$_2$/SiC gate stack is as low as $C_{gate}$=17 pF/cm.

This capacitance is determined only by electrons in the potential well created underneath the ITO electrode as the gate voltage increases from 1.35 to 4 V. Hence, in this transient regime, the inherently bipolar GT-SPED acts as a unipolar device. When the gate voltage increases from 1.35 to 4.0 V, electrons injected from the n-type contact stream to the potential well. This potential well is at a distance of only 0.75 μm from the n-type region (Figs. 1(a) and 1(f)). At $V_{bias}$=2.77 V, the conduction and valence bands are flat in the i-region of the structure (see Figs. 1(e) and 1(f)). Therefore, these electrons stream to the potential well virtually without resistance, i.e., the corresponding resistance $R_{gate}$ is lower than $R_{p-n}$. As a result, the characteristic response time does not exceed $R_{gate}C_{gate} \lesssim 170$ ps. In the fabricated devices, the response time can be slightly higher due to the series resistances of the n-type contact, gate electrode, and the external electronic circuit. Thus, the SPEL of the GT-SPED can be switched on and off almost immediately, whereas the contrast between the ON and OFF states is as high as 60.

To conclude, we have proposed and numerically demonstrated a concept of a gate-tunable single-photon emitting diode based on color centers in wide-bandgap semiconductors. It is based on a lateral p-i-n diode with an optically transparent gate electrode on top of the i-region. When no voltage is applied to the gate, the GT-SPED acts as a conventional single-photon emitting diode, while by applying a voltage to the gate, it is possible to dynamically control the SPEL rate of the color center underneath the gate in a wide range at high speed. We have shown that the SPEL rate of the $Si_C$-SF defect in the 4H-SiC GT-SPED exceeds 40 Mcps at a current density of 30 A/cm$^2$. The SPEL can be switched on and off by applying a gate voltage of less than 5 V. The contrast of the SPEL rates in the ON and OFF states is as high as 60, and the switching time is less than 0.2 ns. These findings provide a solid foundation for the development of efficient and tunable electrically driven single-photon sources based on color center in silicon carbide and other wide-bandgap semiconductors for various quantum information applications.

**Acknowledgments**

The work is supported by the Russian Science Foundation (17-79-20421).

# Gate-tunable single-photon electroluminescence of color centers in silicon carbide

Igor A. Khramtsov and Dmitry Yu. Fedyanin*

*Laboratory of Nanooptics and Plasmonics, Moscow Institute of Physics and Technology, 141700 Dolgoprudny, Russian Federation*

*E-mail: dmitry.fedyanin@phystech.edu

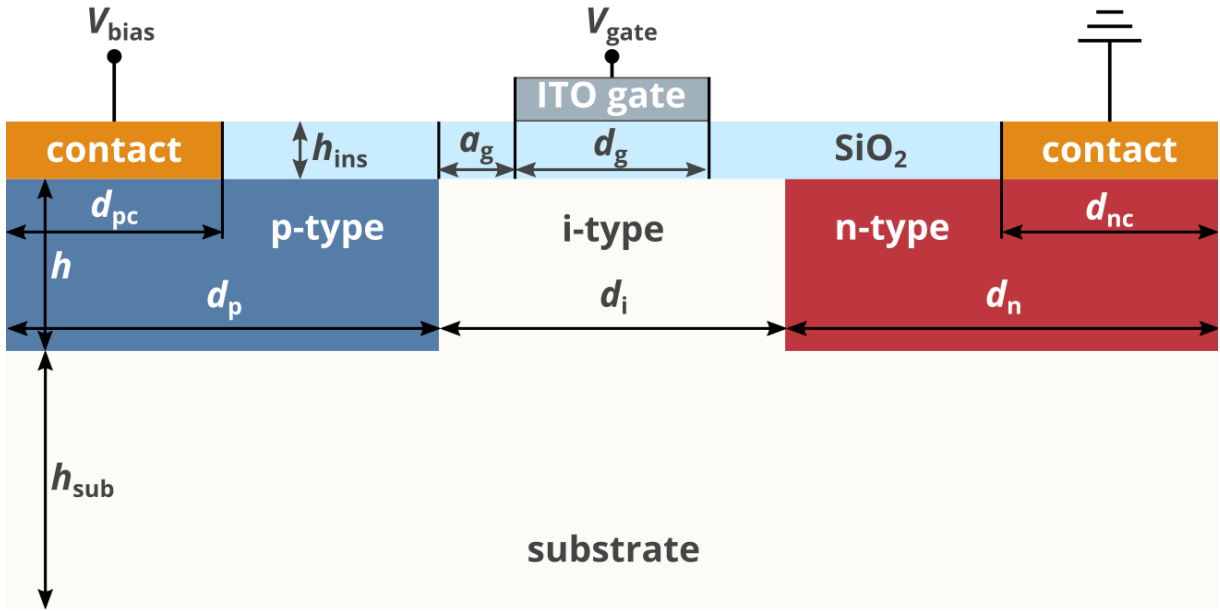

**Figure S1.** Schematics of the numerically simulated 4H-SiC diode with a gate electrode. $d_p = d_n = 3$ μm, $d_i = 2.5$ μm, $d_{pc} = d_{nc} = 1.0$ μm, $a_g = 0.75$ μm, $d_g = 1$ μm, $h = 0.5$ μm, $h_{ins} = 20$ nm, $h_{sub} = 10.0$ μm.

**Table S1.** Main material parameters of 4H-SiC used in the numerical simulations.

| | |
|---|---|
| **Energy bandgap at *T*=300 K, eV** | 3.23 [1] |
| **Dielectric constant** | 9.78 [2] |
| **Density of states effective electron mass** | $0.77m_0$ [3] |
| **Density of states effective hole mass** | $0.91m_0$ [4] |
| **Acceptor concentration in the p-type region, cm$^{-3}$** | $10^{18}$ |
| **Acceptor compensation ratio** | 1% |
| **Acceptor ionization energy, eV** | 0.20 (Al) [5] |
| **Donor concentration in the n-type region, cm$^{-3}$** | $10^{18}$ |
| **Donor compensation ratio** | 1% |
| **Donor ionization energy, eV** | 0.06 (N) [6] |
| **Electron mobility in the p-type region, cm$^2$/Vs** | 500 [7,8] |
| **Hole mobility in the p-type region, cm$^2$/Vs** | 140 [7,8] |
| **Electron mobility in the i-type region, cm$^2$/Vs** | 900 [7,8] |
| **Hole mobility in the i-type region, cm$^2$/Vs** | 140 [7,8] |
| **Electron mobility in the n-type region, cm$^2$/Vs** | 300 [7,8] |
| **Hole mobility in the n-type region, cm$^2$/Vs** | 120 [7,8] |
| **Electron saturation velocity, cm/s** | $2.2 \times 10^7$ [9] |
| **Hole saturation velocity, cm/s** | $1.5 \times 10^7$ [1] |
| **Electron and hole recombination lifetimes in the p-type region, ns** | 17[a] |
| **Electron and hole recombination lifetimes in the i-type region, ns** | 20[a] |
| **Electron and hole recombination lifetimes in the n-type region, ns** | 17[a] |

a) The recombination lifetimes for electrons and holes are calculated using the Scharfetter relation [10]:

$$\tau = \frac{\tau_0}{1 + \frac{N_D + N_A}{N_{ref}}}$$

where $N_D$ and $N_A$ are the concentrations of donors and acceptors, respectively, $N_{ref}$ is estimated to be $5 \times 10^{18}$ cm$^{-3}$ [11] and $\tau_0 = 20$ ns [11].

**Table S2.** Parameters of the Si$_\text{C}$-SF center in 4H-SiC.

| | |
|---|---|
| **Single-photon electroluminescence rate $R$** | $R = \left[ \left( \dfrac{1}{c_\text{n} n} + \dfrac{1}{c_\text{p} p} \right) \left( 1 + \dfrac{\tau_\text{r}}{\tau_\text{nr}} \right) + \tau_\text{r} \left( 1 + \dfrac{\tau_\text{s}}{\tau_\text{nr}} \right) \right]^{-1}$, where $n$ and $p$ are the electron and hole densities in the vicinity of the color center, respectively. The other parameters are listed below. |
| **Radiative lifetime of the excited state $\tau_\text{r}$** | 3.6 ns [12,13] |
| **Nonradiative lifetime of the excited state $\tau_\text{nr}$** | 36 ns [12,13] |
| **Lifetime of the shelving state $\tau_\text{s}$** | 33 ns [13] |
| **Electron capture constant $c_\text{n}$** | $1.1 \times 10^{-7}$ cm$^3$s$^{-1}$ [13] |
| **Hole capture constant $c_\text{p}$** | $5.4 \times 10^{-8}$ cm$^3$s$^{-1}$ [13] |